\documentclass[a4paper,10pt] {article}

\usepackage{fcds}
\usepackage{graphicx}
\usepackage{float}
\usepackage[labelsep=period]{caption}
\usepackage{enumitem}
\usepackage[utf8]{inputenc}
\usepackage{titlesec}
\usepackage[table,xcdraw]{xcolor} 
\usepackage{multirow}
\usepackage{multicol}
\titlelabel{\thetitle.\quad}
\usepackage[top=5.5cm, bottom=5cm, left=4cm, right=4cm]{geometry}
\graphicspath{ {./pics/} }

\date{}

\title{Comparative review of selected Internet communication protocols}
\author{Łukasz Kamiński
\and 
Maciej Kozłowski,
\and 
Daniel Sporysz, 
\and 
Katarzyna Wolska, 
\and
Patryk Zaniewski, 
\and
Radosław Roszczyk}
\date{April 2022}

\begin{document}

\maketitle

\begin{abstract}

With a large variety of communication methods and protocols, many software architects face the problem of choosing the best way for services to share information. For communication technology to be functional and practical, it should enable developers to define a complete set of CRUD methods for the processed data. The research team compared the most commonly used data transfer protocols and concepts in this paper: REST, WebSocket, gRPC GraphQL and SOAP. To do that, a set of web servers was implemented in Python, each using one of the examined technologies. Then, the team performed an automated benchmark measuring time and data transfer overhead for a set of defined operations: creating an entity, retrieving a list of 100 entities and fetching details of one entity. Tests were designed to avoid the results being interfered by database connection or docker-compose environment characteristics. The research team has concluded that gRPC was the most efficient and reliable data transfer method. On the other hand, GraphQL turned out to be the slowest communication method of all. Moreover, its server and client libraries caused the most problems with proper usage in a web server. SOAP did not participate in benchmarking due to limited compatibility with Python and a lack of popularity in modern web solutions.
    
\end{abstract}


\section{Introduction}

In modern world of technology there are many tools that developer can choose in order to exchange data between web client such as web page and a server. These tools wary in speed and the amount of sent data. Therefore, it is not an easy task to make a decision which tool to use in a project. Our team recognized that problem and decided to check which of common web protocol is the fastest and the most memory efficient. We decided to test gRPC, REST architectural style, the WebSockets Protocol and GraphQL. We skipped SOAP protocol because of its lack of popularity in modern web applications.

This paper is organized as follows. Next subsections briefly describe tested protocols. Section two presents related articles. Section three provides the description of testing methods and testing environment. Section four presents test results. Finally, section five discusses them ad provides concluding remarks.

\subsection{REST}
REST (Representational State Transfer) is an architectural style for distributed systems that was first described by Roy T. Fielding in 2000. It is defined by several constraints that will be shortly related.

The idea of this approach is based on widely known client-server architecture. REST allows client and server to work independently by separating user interface on the client side and business logic with data storage on the server side. Thanks to that, REST achieves the portability of the user interface across multiple, different platforms.

Another constraint of this architecture is the fact, that the communication between the system parts must be stateless. It means that each request from the client to the server must contain all of the information necessary to understand and complete the request. Because of that, only the client side is responsible for keeping the session state.

Next principle is the cacheability of responses. It states that every response must inform the client if its cacheable or noncacheable, because if it is, the client is able to reuse that data for later, similar requests.

After that, there is a uniform interface constraint, which suggest that by "applying the software engineering principle of generality to the component interface, the overall system architecture is simplified and the visibility of interactions is improved". This rule also describes some more requirements for REST interface to work properly, such as the use of hyperlinks or unique identification or each resource.

REST architecture also specifies that it requires the system to be layered. The layered system style allows an architecture to be composed of hierarchical layers by constraining component behavior.

The last constraint is Code-On-Demand. It means that REST can also allow client to download and execute code in form of applets or scripts.

In REST architecture the sample of information is referred as resource. Each resource has its identifier which is the URL that points to the specified piece of information. To manage this data, REST application provide RESTful APIs, where resources are exposed by means of endpoints. The endpoints, in general are the URLs that identifies resources.

REST is strictly related to HTTP protocol. It is caused by the fact that to interact with resources in a REST system, we use 4 basic HTTP methods. These are:
\begin{itemize}
    \item GET -- used to retrieve list of resources or specific resource by its ID
    \item POST -- used to create new resource
    \item PUT -- used to update resource by its ID
    \item DELETE -- used to delete resource by its ID
\end{itemize}

There are also some data serialization formats that are used in REST communication. The most popular are JSON and XML. Using specified in system format, we send resources to create or update and also the server provides us information of resources we requested in that format. 

Summing up, usage of REST architecture can result in better performance and scalability in simple, lightweight applications. \cite{http_rfc} \cite{rest_roy}

\subsection{The WebSocket Protocol}
The WebSocket Protocol is a protocol which enables two-way communication. This term means that server doesn't have to wait for client to make a request and can send  data independently to it. This is achieved by using a single TCP connection which is established between client and server. \cite{rfc6455}

The protocol consists of two parts. The first one is a handshake. In order to make it client sends a HTTP request. This request must have several headers set (Connection, Upgrade). Example client request:
\begin{verbatim}
GET /chat HTTP/1.1
Host: server.example.com
Upgrade: websocket
Connection: Upgrade
Sec-WebSocket-Key: dGhlIHNhbXBsZSBub25jZQ==
Origin: http://example.com
Sec-WebSocket-Protocol: chat, superchat
Sec-WebSocket-Version: 13
\end{verbatim}
Server replies with its own handshake response which has HTTP code set to 101. Example response:
\begin{verbatim}
HTTP/1.1 101 Switching Protocols
Upgrade: websocket
Connection: Upgrade
Sec-WebSocket-Accept: s3pPLMBiTxaQ9kYGzzhZRbK+xOo=
Sec-WebSocket-Protocol: chat
\end{verbatim}

When the handshake is successfully finished further communication is done via single TCP connection using The Websocket Protocol. Client and server can send data independently in chunks named as messages. Messages are made of frames which are transferred across sides.

In order to close connection a special frame is sent by one of the peers. This frame indicates the will of closure. When other peer receives it, it responds with a Close frame. Thanks to that the first peer can safely close the connection knowing that there will be no more data transferred. 

\subsection{gRPC}

gRPC (Google Remote Procedure Call) is an open source system that realizes remote procedure call concept. It was developed in 2015 by Google as a successor of previous RPC system called Stubby. 

The power of gRPC comes from many things but there are two which are major. It is HTTP/2 communication protocol and Protocol Buffers.

HTTP/2 as the newest version of popular HTTP protocol provides a lot of useful features to this type of communication. Thus, gRPC provides the mechanisms of deadlines, timeouts, terminations and cancelling calls. Moreover, thanks to this protocol, there are 4 types of communication that can be implemented \cite{grpc_doc}:
\begin{itemize}
    \item Unary Call - the simplest type of communication where there is only one for one request
    \item Client Stream - in some way very similar to Unary Call but here client sends stream of messages instead one
    \item Server Stream - the opposition of Client Stream where client sends only one request and expects the stream of messages
    \item Bidirectional Stream - combination of two previous methods where two sides of stream messages at the same time
\end{itemize}

Protocol Buffers is a cross-platform mechanism developed by Google which allows to define data structures and interfaces in *.proto files. Then, developed file can be used to generate a file in targeted language (i.e. Java, Python, Go). The benefit of this is that one Protobuf file can generate an API for different language that is implementing server side, and other which is implementing client side. What is more, Protocol Buffer is a binary format what means that is faster and smaller than other formats (i.e. JSON) \cite{proto_doc}.

\subsection{SOAP}

Simple Object Access Protocol (SOAP) defines a mechanism for exchanging structured and typed information in a decentralized, distributed environment. It it possible to express application semantics with packaging model and encoding rules for data within modules. It is important to note that SOAP does not define any application semantics itself but a mechanism for expressing it so it has many use cases ranging from messaging systems to RPC.

Each SOAP message consists of envelope, encoding rules and RPC representation. SOAP envelope represents a message framework describing what is a message, who is its receiver and which data is mandatory or optional. SOAP encoding rules explain a serialization and de-serialization mechanism for application data types. SOAP RPC representation constructs remote procedure calls and responses.

SOAP data model was designed with language-independent abstraction for data types. Two data types are supported: simple XSD and compound. For example, simple XSD types define int, boolean and strings. Compound types are used to exchange structures and arrays. Simple Object Access Protocol is XML based and each message serializes into XML file.

\subsection{GraphQL}

GraphQL is an open-source solution for querying and mutating data via remote API. The solution was first developed internally by Facebook, but it became publicly released in 2015. Now it is developed under Linux Foundation.

The core concept of GraphQL relies on an extensive query language that enables API consumers to not only apply filters, but also choose fields of the returned structure. This approach is more flexible than REST APIs exposing a set of endpoints, each returning the consistent set of data. The responsibility for defining the presence of individual fields in a returned structure is moved from an API server to an API consumer, which results in an optimization of data traffic and an ability to eliminate API versioning. GraphQL user HTTP protocol for network communication.

GraphQL lets API developers declare an interface for querying data on SQL and NoSQL database solutions. Since GraphQL allows for nested structures, a NoSQL database (e.g. MongoDB) is a great fit for the technology. GraphQL also improves development flexibility by providing client-side and server-side API libraries for many popular programming languages (e.g. Python used in this research).

\section{Related work}

We have searched for similar works, that intended to compare REST, gRPC, Websockets and GraphQL protocols, but we have found none that would compare all of them in terms of performance. However, there are studies that set some of those protocols side by side, for example, contrasting the time it takes to implement a query or analysing an energy costs of specified protocols.

Many of the found papers focus on testing REST and GraphQL protocols. In the first work, authors lead a performance comparative study with the stated two technologies. They analysed three applications and each of them was developed using two protocols -- REST and GraphQL. The researched was based on measuring the response time and the average transfer rate between the requests. The study describes, that in two applications migrating from REST to GraphQL resulted in higher performance in the aspects of average number of requests per second and transfer rate of data. However, above workloads of 3000 requests, REST protocol performed better than GraphQL. \cite{rest_graphql}

Next study compares REST and GraphQL in terms of convenience of use. It answers the question "How much time do developers spend when implementing queries in REST and GraphQL?". Results of the experiment, which was described in that paper, show that implementing remote service queries was more quick, when using GraphQL technology than REST technology. Implementing GraphQL queries took less effort compared to REST, among participants with no or little knowledge of those protocols, as well as among experienced developers. \cite{rest_graphql_convenience}

Next paper examines the performance of REST and GraphQL protocols. A series of experiments was also executed that showed that the choice of better protocol for given purpose depends highly on type of query and requested data. With the same amount of used data, REST protocol is obvious better choice as it stands out with better response times, CPU and memory consumption. However, in test cases where REST was forced to under-fetch data, GraphQL performed better in terms of response time and memory usage but still needed more of the CPU resources. Overall the result showed that GraphQL should be considered when under- and over-fetching or request smaller subsets of data is likely to happen and REST protocol is the best option in other, more basic cases. \cite{frigaard2022graphql}

The following work contrasts REST, SOAP, Socket and gRPC in computation offloading of mobile applications and it analysis energy costs of those protocols. The experiment carried out in that study was about evaluating the energy consumption of stated protocols using algorithms of different complexities and different input sizes and types. Results show that, when executing more simple algorithms with small input data, local execution is way more economic. Regarding remote execution, the best option is REST architecture, that is followed by Socket. The paper also states, that computation offloading can be responsible of saving up to 10 times as much energy when compared to local execution. \cite{energy_cost}

After that, there is also a paper that analyzes the efficiency of REST and gRPC protocols in microservice-based ecosystems. In order to perform tests, the authors created implementations of REST and gRPC services which were developed using .NET 5 platform. The main parameter that was tested, was the response time of the performed operations. The explored communication tasks was based on: text cloning, fetching maximum value of an integer, fetching an array of consecutive integers, fetching a text file and downloading a PDF file. Each test was performed with the use of both encrypted and not encrypted data. For most of the tasks that was tested with not encrypted data, the results was better for the REST protocol. Only in the large file transfer gRPC performed better. On the other hand, with the usage of encrypted data, both protocols got similar results. REST performed better during transmission of numerical data and gRPC was faster for file transfer operations. \cite{efficiency_rest_grpc}

Next work describes performance comparison between GraphQL, REST and SOAP protocols. The method used to evaluate the differences between those three protocols, was based on data fetching operation. For the test purpose, authors created systems for each protocol using .NET technologies. There were several test cases that included fetching elements from database, as well as using simple and more complex joins. These operations was tested for a single row and also for 100 rows. For the analysis, the authors also used two types of databases: MySQL and MongoDB. The experiment was performed using two computers -- server and client -- that were connected through a local network. The results showed that GraphQL was characterized by the worst performance in all test cases. Another conclusion is the fact that the packet size is the largest with the usage of SOAP protocol. It is caused by the XML format used in SOAP message passing. The performance results of the REST protocol were the best among tested technologies. \cite{performance_graph_rest_soap}

Another study compares the performance of REST API, GraphQL and gRPC. For the research, the authors developed three applications that contained the same functionalities, but with the use of different protocols. The systems was created using .NET 5 platform. The experiments measured the execution time, performance and volume of the data, that was processed during display and adding operations. The exact testing methods relied on fetching small, medium and large amount of data, as well as inserting new data. The result showed, that the best protocol in terms of performance is REST. However, considering the smallest data package size, the better option is gRPC protocol. Overall, the choice of selecting one specific protocol for a given task is very complicated. During selecting a technology, several factors should be considered such as: data size, system performance and number of users. \cite{sliwa_panczyk}

The following work contrasts WebSocket and HTTP protocol performance. For the testing purposes the authors used machines working in the same LAN network and developed a special application. The methods used for the analysis depended on sending and receiving texts of the length of 100 characters. The main conclusion from the research is the fact that with the transmission of over 100 data copies using the WebSocket protocol can result in over 100 times better performance over HTTP protocol. It was also proved that the usage of WebSockets is a good way to transfer a big number of small data packages in the period of one second, because in other scenarios basic HTTP requests are better option. The authors also noticed that the TLS encryption has no effect on the performance of both protocols. \cite{lasocha_badurowicz}

After that, there is also a paper that compares the performance of web services using Symfony, Spring,
and Rails technologies. Using each of the frameworks, REST and SOAP application were developed. In research the authors focused on measuring the request execution time. The tests included select, insert and update operations. Results showed that performance of the REST and SOAP protocols highly depends on technology in which the application was developed. \cite{lubartowicz2020performance}

Last study describes performance and usage comparison between REST and SOAP web services. The test results showed the fact, that REST to outperform SOAP in terms of bandwidth usage and message processing performance. Authors also stated, that REST is a good option in basic, most common cases, while SOAP should be considered if particular functionality, such as security options, is required. REST is also a more simple, easier to develop technology than SOAP protocol. \cite{makkonen2017performance}

\section{Measurement methods}

In this paper, we discuss the differences in features and limitations of communication protocols for web services to help engineers and architects with choosing a protocol that suits their needs the most. Although the knowledge of their functionalities is often sufficient enough to select one, some solutions are required to handle large volume of requests and we are not able to make an informed decision without an insight into protocol performance. We would like to perform benchmarks to find the fastest protocols and protocols with the least network data footprint. 

Nowadays most web services are deployed in a containerized environment. We are interested in knowing how much performance is allocated for virtualization thus we will perform benchmarks comparing performance of applications deployed in a bare metal environment and a virtualized environment.

\subsection{Performance comparison}

For each protocol we implemented a web service providing an interface to CRUD operations on a database using libraries listed in the table \ref{table:libraries}. Main factors for choosing these libraries are popularity and deployment web frameworks. With popular projects we are less likely to encounter unusual performance issues and bugs. Also notice how Flask is used across three out of four implementations. Using the same web framework should provide results that are better comparable.

To evaluate the performance of protocols we measured a total time elapsed between making a request and fully receiving a response. To achieve this, we implemented clients in Python for each service using libraries from table \ref{table:client_libraries}. The clients, the servers and a database were hosted on the same machine. Although this architecture could result in them competing for resources, we believe the benefits of reduced network delay outweigh a solution in which we could deploy these projects on separate host machines.

We used Mongodb 5.0.5 database and Steam Games DataSet \cite{80000_Steam_Games_DataSet}. NoSQL database was chosen because we want to minimize the impact of database operations on our measurements and NoSQL databases are generally faster than SQL databases \cite{inproceedingsSqlNoSqlPerf}.

Three scenarios were considered in test cases: inserting an entry, fetching an entry and fetching many entries. First we measured how performant protocols are with inserting a single game entry into a database. The test was performed on an empty database with indexing turned off. Before proceeding to another protocol, the database was restored to an initial state to avoid performance degradation over time. The same game entry data was used for testing of every service. Then indexing was turned on and the database was populated with 100 entries from Steam Game DataSet. We used just 100 entries to keep a table index short and make queries to database as quick as possible. The other two test cases aim to measure how protocols behave with small and large outbound data transfers. The second and the third test cases involve fetching a single game entry and 100 game entries at once respectively. Same as for the first test, we ensured that the same data was used to test each protocol.

Before any benchmark took place, we generated a load on each service to address the issue of cold start.

\begin{table}[]
\centering
\caption{API libraries and versions}
\begin{tabular}{|l|p{7cm}|lll}
\cline{1-2}
\cellcolor[HTML]{EFEFEF}Project & \cellcolor[HTML]{EFEFEF}Used libraries and versions &  &  &  \\ \cline{1-2}
REST                            & Flask 2.0.3, requests 2.27.1, pymongo 4.0.2  &  &  &  \\ \cline{1-2}
GraphQL                         & Flask 2.0.3, graphql-core 2.3.2, graphql-relay 2.0.1, graphql-server-core 2.0.0, graphene 2.1.9, graphene-mongo 0.2.13 &  &  &  \\ \cline{1-2}
WebSockets                      & Flask 2.0.3, simple-websocket 0.5.1, pymongo 4.0.2 &  &  &  \\ \cline{1-2}
gRPC                            & pymongo 4.0.2, grpcio 1.45.0, grpcio-tools 1.45.0 &  &  &  \\ \cline{1-2}
\end{tabular}
\label{table:libraries}
\end{table}

\begin{table}[]
\centering
\caption{Client libraries and versions}
\begin{tabular}{|l|p{7cm}|lll}
\cline{1-2}
\cellcolor[HTML]{EFEFEF}Project & \cellcolor[HTML]{EFEFEF}Used client libraries and versions &  &  &  \\ \cline{1-2}
REST client                            & requests 2.27.1  &  &  &  \\ \cline{1-2}
GraphQL client                         & python-graphql-client 0.4.3 &  &  &  \\ \cline{1-2}
WebSockets client                      & simple-websocket 0.5.1 &  &  &  \\ \cline{1-2}
gRPC client                            & grpc 1.45.0 &  &  &  \\ \cline{1-2}
\end{tabular}
\label{table:client_libraries}
\end{table}

\subsection{Network load comparison}

To evaluate the impact on network we measured how much data is sent and received to perform an operations on a web service. We captured and recorded every network packet that was sent and received during a single request. Then the data consumption was evaluated by assessing the length of every packet and summing them up. Two operations were tested against data consummations: add one entry and get one entry. All tests were performed on the same entry data.

\subsection{Test environment}

All benchmarks were performed on a platform with Windows 10 OS with hardware configuration specified in a table \ref{table:specs}. To minimize the impact of a real time system on the results of benchmarks we ensured no background tasks were running on a test machine. To test the performance of a vitalized environment we used Docker Desktop 4.6.1 as a virtualization platform. The same Python interpreter in version 3.9.12 was used across all implementations to make results better comparable.

\begin{table}[]
\caption{Hardware configuration}
\begin{center}
\begin{tabular}{|l|l|lll}
\cline{1-2}
\cellcolor[HTML]{EFEFEF}Component & \cellcolor[HTML]{EFEFEF}Model &  &  &  \\ \cline{1-2}
CPU                               & AMD Ryzen 5600X               &  &  &  \\ \cline{1-2}
OS                                & Windows 10 Pro                &  &  &  \\ \cline{1-2}
RAM                               & DDR4 16GB 3200Mhz             &  &  &  \\ \cline{1-2}
\end{tabular}
\end{center}
\label{table:specs}
\end{table}

\newpage

\section{Results}
This section presents the results of conducted tests. We divided them into several subsections. First subsection is covering the performance comparison. It shows the tests of inserting one element to database, getting one element and getting a hundred elements. We ran tests on native OS and on Docker. Second subsection presents how much data needs to be transferred between client and server in order to achieve those operations.
\subsection{Performance comparison}

\subsubsection{Inserting one element to database}
\begin{figure}[htbp!]
    \centering
    \includegraphics[scale=0.5]{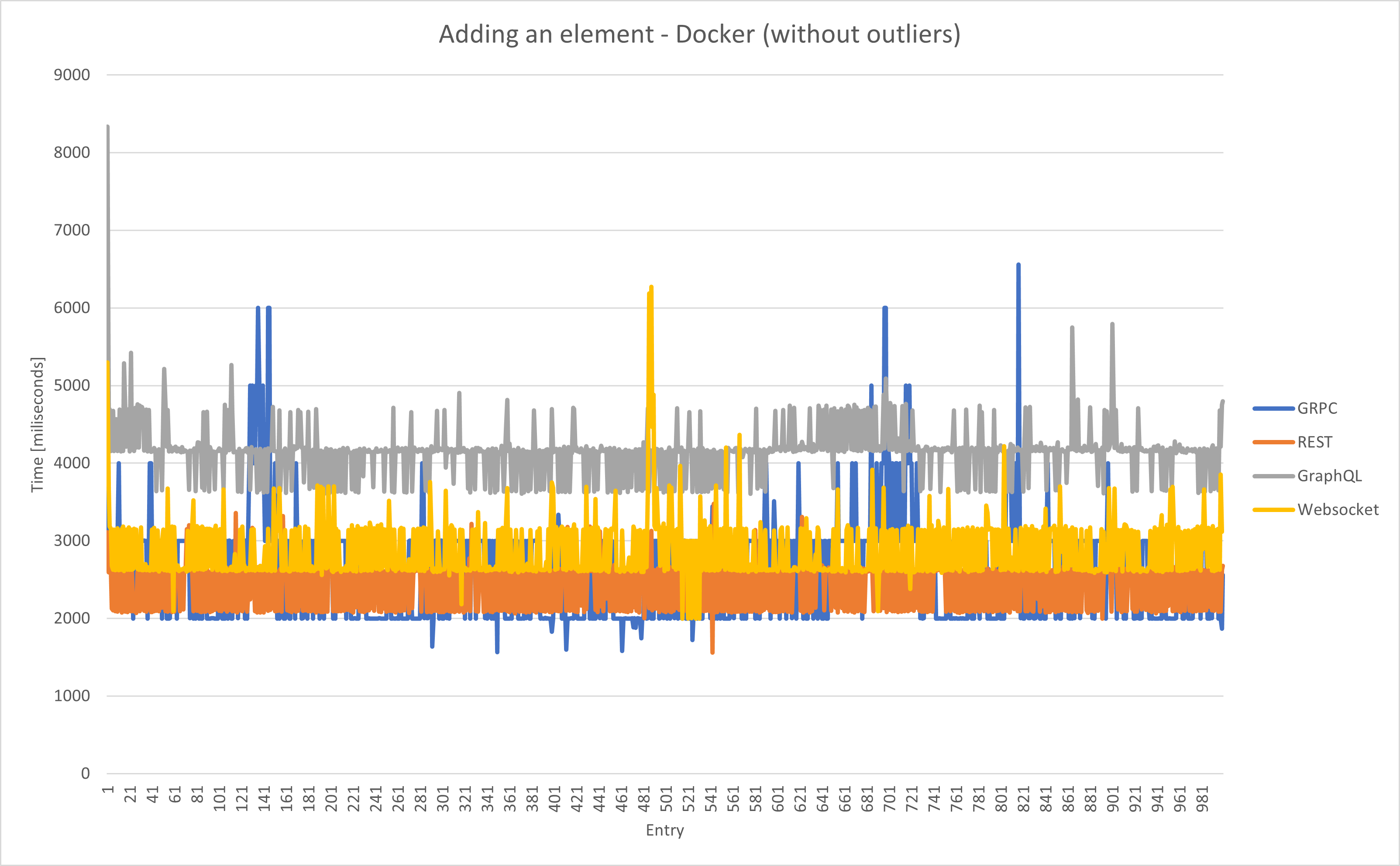}
    \caption{Inserting one element to database on Docker}
    \label{fig:insert_one_docker}
\end{figure}

Figure \ref{fig:insert_one_docker} shows times needed to insert one element to database. We decided to remove outliers from the chart in order to remain readability. We achieved lowest mean times on Docker when using REST style. Surprisingly, gRPC turned out to be slower on this platform.

\begin{figure}[htbp!]
    \centering
    \includegraphics[scale=0.5]{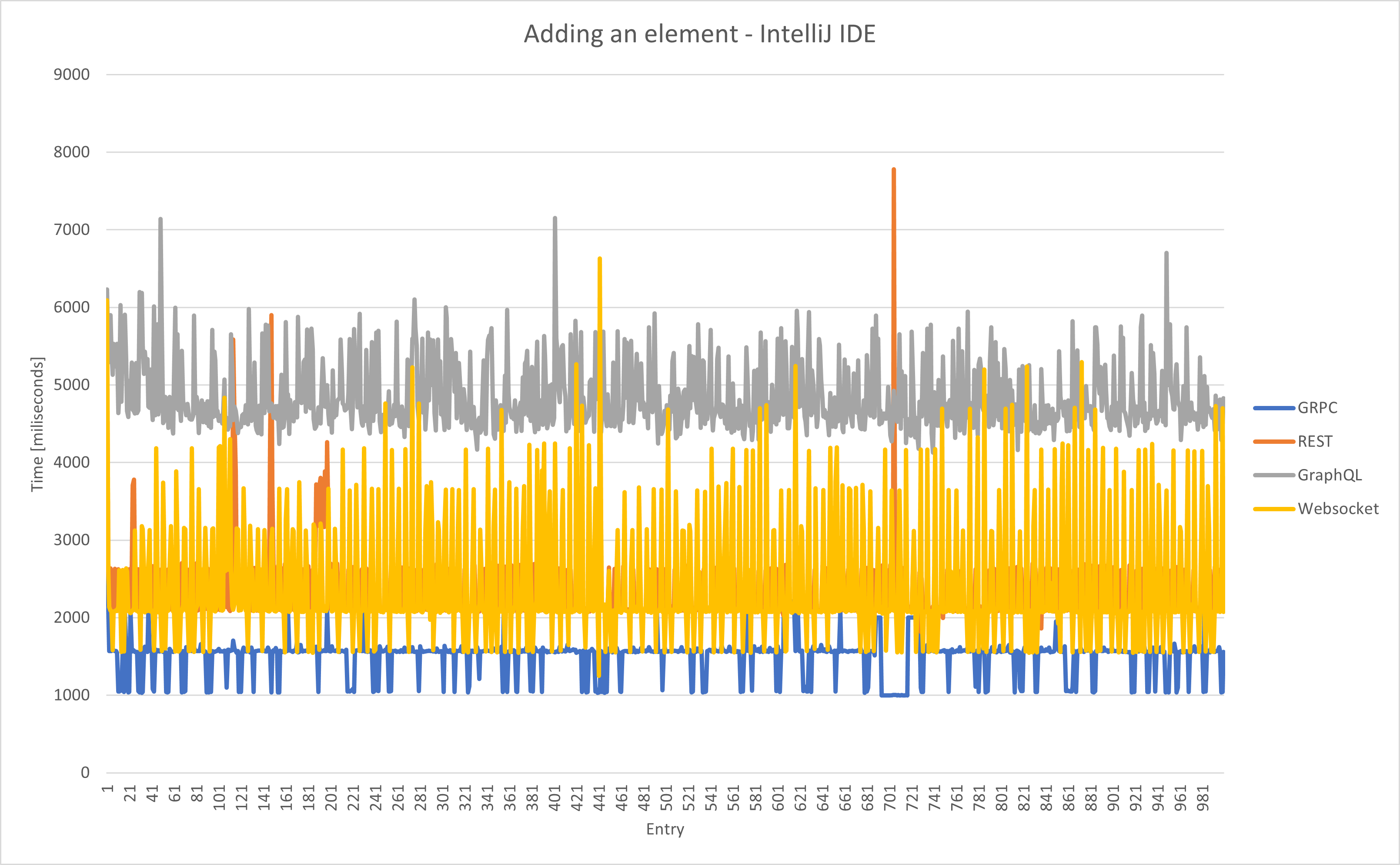}
    \caption{Inserting one element to database on Windows}
    \label{fig:insert_one_native}
\end{figure}

Worth mentioning is the comparison between running tests via Docker and direct on Windows. Figure \ref{fig:insert_one_native} shows lower times for three protocols. This means that the inserting operation is faster on native OS than on Docker. This time gRPC protocol was the fastest of all. Only GraphQL achieved worse performance on native OS than on Docker. What is more, Docker turned out to be less stable than Windows. We noticed several test which took abnormally long. It is especially visible in maximum value for WebSockets tests. The value is more than ten times bigger than on similar test on Windows.

Table \ref{tab:mean_insert_one} presents collected values for Docker and Windows tests.

\begin{table}[htbp!]
    \centering
    \caption{Times measured for inserting one value to database}
    \begin{tabular}{|c|l|r|r|}
        \hline
         \multicolumn{2}{|c|}{\cellcolor[HTML]{EFEFEF}}  & \cellcolor[HTML]{EFEFEF}Docker [$\mu s$] & \cellcolor[HTML]{EFEFEF}Windows [$\mu s$] \\
        \hline
        \cellcolor[HTML]{EFEFEF} & \cellcolor[HTML]{EFEFEF}mean & 2425 & 2278 \\
        \cline{2-4}
        \cellcolor[HTML]{EFEFEF} & \cellcolor[HTML]{EFEFEF}min & 1563 & 1818 \\
        \cline{2-4}
        \cellcolor[HTML]{EFEFEF} & \cellcolor[HTML]{EFEFEF}max & 3479 & 7779 \\
        \cline{2-4}
        \multirow{-4}{*}{\cellcolor[HTML]{EFEFEF}REST} & \cellcolor[HTML]{EFEFEF}$\sigma$ & 280 & 395 \\
        \hline
        \cellcolor[HTML]{EFEFEF} & \cellcolor[HTML]{EFEFEF}mean & 2695 & 1512 \\
        \cline{2-4}
        \cellcolor[HTML]{EFEFEF} & \cellcolor[HTML]{EFEFEF}min & 1565 & 999 \\
        \cline{2-4}
        \cellcolor[HTML]{EFEFEF} & \cellcolor[HTML]{EFEFEF}max & 6560 & 4822 \\
        \cline{2-4}
        \multirow{-4}{*}{\cellcolor[HTML]{EFEFEF}gRPC} & \cellcolor[HTML]{EFEFEF}$\sigma$ & 667 & 253 \\
        \hline
        \cellcolor[HTML]{EFEFEF} & \cellcolor[HTML]{EFEFEF}mean & 4161 & 4834 \\
        \cline{2-4}
        \cellcolor[HTML]{EFEFEF} & \cellcolor[HTML]{EFEFEF}min & 3605 & 4128 \\
        \cline{2-4}
        \cellcolor[HTML]{EFEFEF} & \cellcolor[HTML]{EFEFEF}max & 8337 & 7151 \\
        \cline{2-4}
        \multirow{-4}{*}{\cellcolor[HTML]{EFEFEF}GraphQL} & \cellcolor[HTML]{EFEFEF}$\sigma$ & 318 & 429 \\
        \hline
        \cellcolor[HTML]{EFEFEF} & \cellcolor[HTML]{EFEFEF}mean & 2950 & 2424 \\
        \cline{2-4}
        \cellcolor[HTML]{EFEFEF} & \cellcolor[HTML]{EFEFEF}min & 1999 & 1248 \\
        \cline{2-4}
        \cellcolor[HTML]{EFEFEF} & \cellcolor[HTML]{EFEFEF}max & 90272 & 6632 \\
        \cline{2-4}
        \multirow{-4}{*}{\cellcolor[HTML]{EFEFEF}WebSockets} & \cellcolor[HTML]{EFEFEF}$\sigma$ & 2789 & 775 \\
        \hline
    \end{tabular}
    \label{tab:mean_insert_one}
\end{table}

\newpage 

\subsubsection{Fetching one element}
\begin{figure}[htbp!]
    \centering
    \includegraphics[scale=0.5]{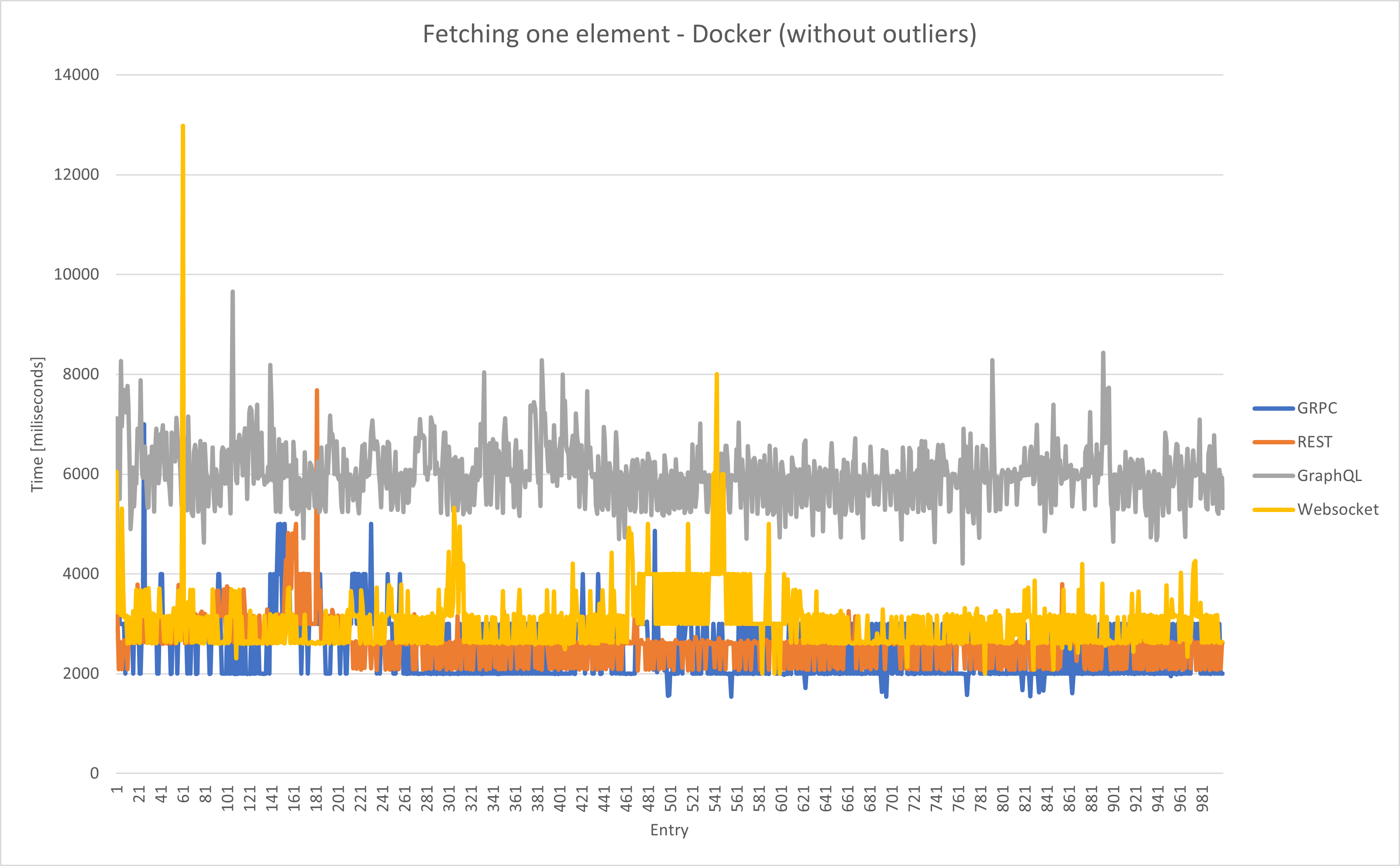}
    \caption{Fetching one element on Docker}
    \label{fig:fetch_one_docker}
\end{figure}

\begin{figure}[htbp!]
    \centering
    \includegraphics[scale=0.5]{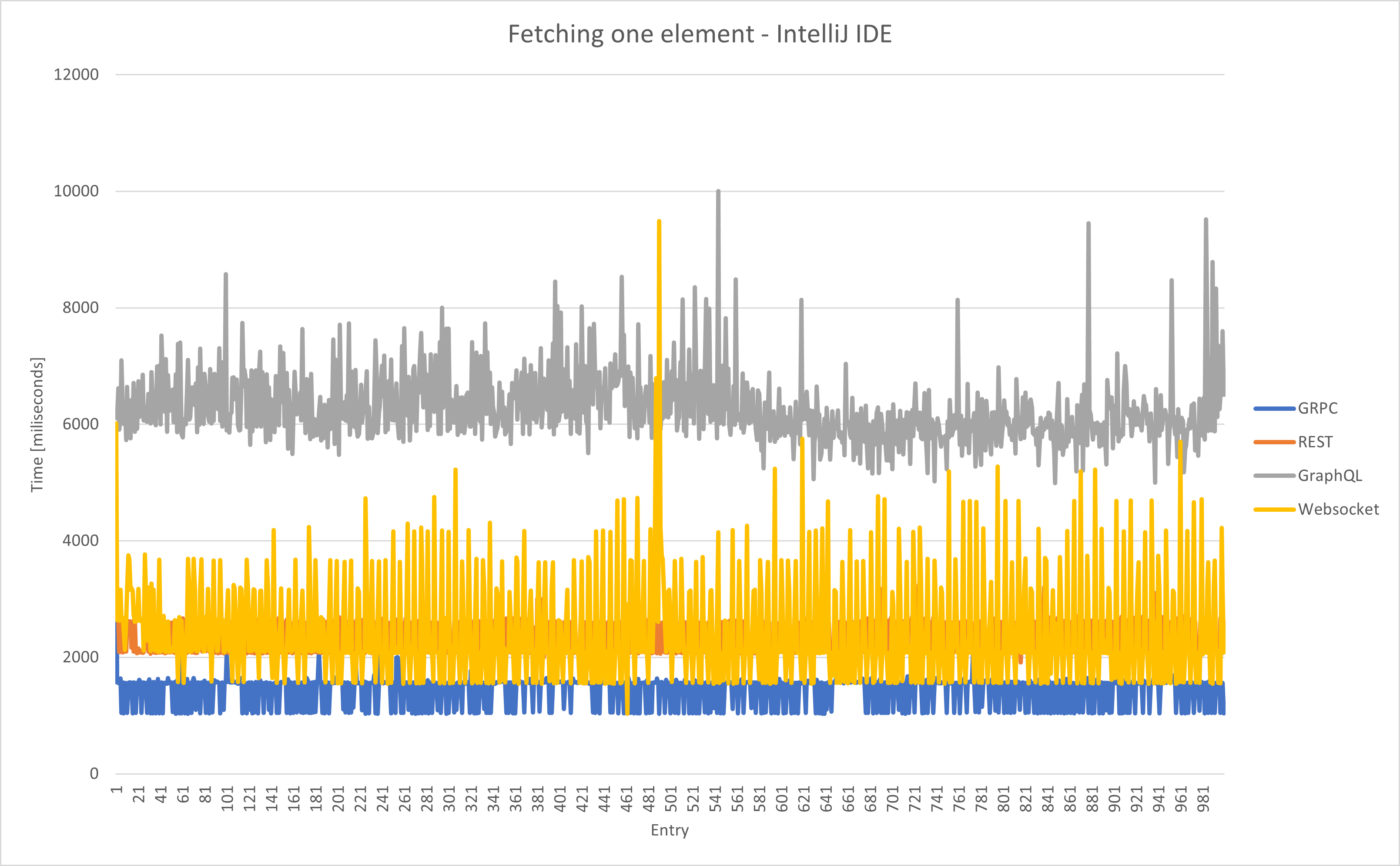}
    \caption{Fetching one element on native OS}
    \label{fig:fetch_one_nativeOS}
\end{figure}

Our second test was to measure times needed to fetch one element from database. Figure \ref{fig:fetch_one_docker} shows result chart of tests done on Docker platform, while figure \ref{fig:fetch_one_nativeOS} presents results from Windows. Once more times measured during tests conducted on native OS were shorter.

GraphQL protocol was the slowest both on Docker and Windows. REST and gRPC achieved very similar results, but gRPC turned out to be the fastest of all. Its speed is especially visible when testing on native OS. WebSockets were the least stable again. Their maximum test value was several times bigger than other protocols. Table \ref{tab:mean_fetch_one} shows test times.

\begin{table}[htbp!]
    \centering
    \caption{Times measured for fetching one value from database}
    \begin{tabular}{|c|l|r|r|}
        \hline
         \multicolumn{2}{|c|}{\cellcolor[HTML]{EFEFEF}}  & \cellcolor[HTML]{EFEFEF}Docker [$\mu s$] & \cellcolor[HTML]{EFEFEF}Windows [$\mu s$] \\
        \hline
        \cellcolor[HTML]{EFEFEF} & \cellcolor[HTML]{EFEFEF}mean & 2631 & 2289 \\
        \cline{2-4}
        \cellcolor[HTML]{EFEFEF} & \cellcolor[HTML]{EFEFEF}min & 2066 & 1837 \\
        \cline{2-4}
        \cellcolor[HTML]{EFEFEF} & \cellcolor[HTML]{EFEFEF}max & 7683 & 5134 \\
        \cline{2-4}
        \multirow{-4}{*}{\cellcolor[HTML]{EFEFEF}REST} & \cellcolor[HTML]{EFEFEF}$\sigma$ & 428 & 280 \\
        \hline
        \cellcolor[HTML]{EFEFEF} & \cellcolor[HTML]{EFEFEF}mean & 2495 & 1456 \\
        \cline{2-4}
        \cellcolor[HTML]{EFEFEF} & \cellcolor[HTML]{EFEFEF}min & 1537 & 1029 \\
        \cline{2-4}
        \cellcolor[HTML]{EFEFEF} & \cellcolor[HTML]{EFEFEF}max & 7001 & 5101 \\
        \cline{2-4}
        \multirow{-4}{*}{\cellcolor[HTML]{EFEFEF}gRPC} & \cellcolor[HTML]{EFEFEF}$\sigma$ & 616 & 258 \\
        \hline
        \cellcolor[HTML]{EFEFEF} & \cellcolor[HTML]{EFEFEF}mean & 6022 & 6271 \\
        \cline{2-4}
        \cellcolor[HTML]{EFEFEF} & \cellcolor[HTML]{EFEFEF}min & 4209 & 4989 \\
        \cline{2-4}
        \cellcolor[HTML]{EFEFEF} & \cellcolor[HTML]{EFEFEF}max & 9656 & 10000 \\
        \cline{2-4}
        \multirow{-4}{*}{\cellcolor[HTML]{EFEFEF}GraphQL} & \cellcolor[HTML]{EFEFEF}$\sigma$ & 588 & 578 \\
        \hline
        \cellcolor[HTML]{EFEFEF} & \cellcolor[HTML]{EFEFEF}mean & 3069 & 2439 \\
        \cline{2-4}
        \cellcolor[HTML]{EFEFEF} & \cellcolor[HTML]{EFEFEF}min & 1999 & 1031 \\
        \cline{2-4}
        \cellcolor[HTML]{EFEFEF} & \cellcolor[HTML]{EFEFEF}max & 30320 & 9491 \\
        \cline{2-4}
        \multirow{-4}{*}{\cellcolor[HTML]{EFEFEF}WebSockets} & \cellcolor[HTML]{EFEFEF}$\sigma$ & 1071 & 843 \\
        \hline
    \end{tabular}
    \label{tab:mean_fetch_one}
\end{table}

\newpage

\subsubsection{Fetching a hundred elements}
\begin{figure}[hbtp!]
    \centering
    \includegraphics[scale=0.5]{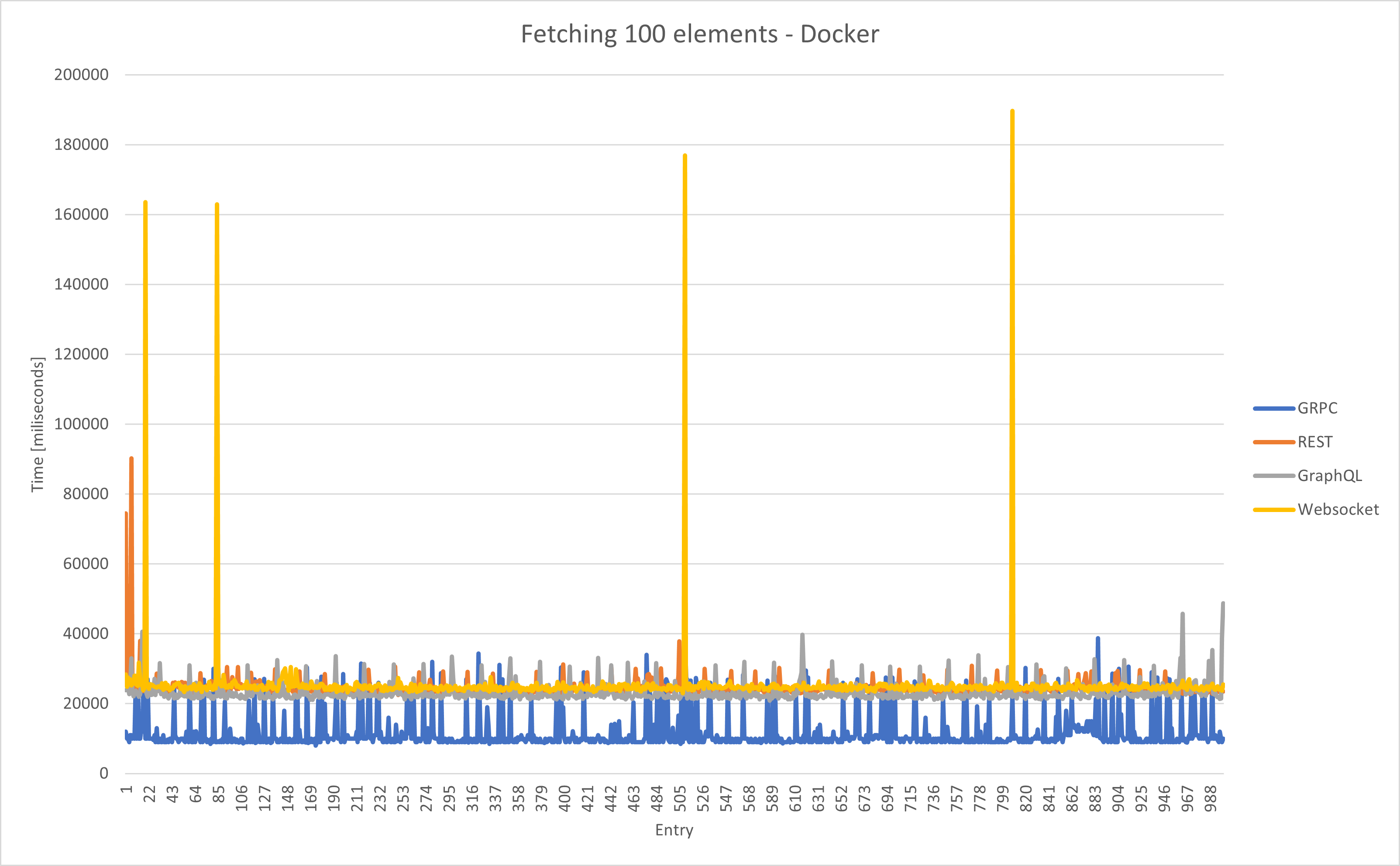}
    \caption{Fetching a hundred elements on Docker}
    \label{fig:fetch_100_docker}
\end{figure}

\begin{figure}[htbp!]
    \centering
    \includegraphics[scale=0.5]{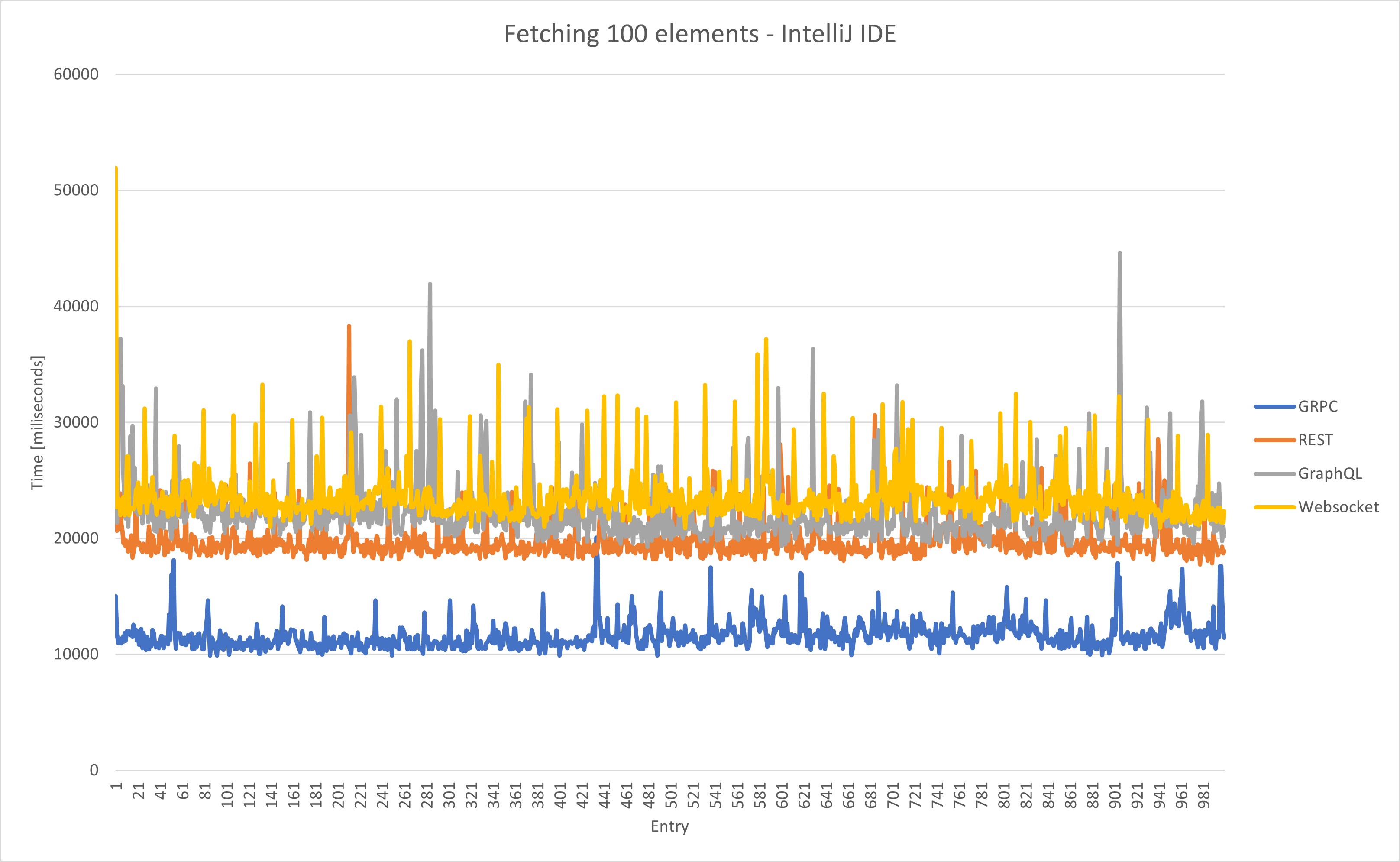}
    \caption{Fetching a hundred elements on native OS}
    \label{fig:fetch_100_nativeOS}
\end{figure}

The last conducted test measured how long it takes to fetch a hundred elements from database. Figure \ref{fig:fetch_100_docker} present results from Docker platform, figure \ref{fig:fetch_100_nativeOS} shows results from Windows. 

This time the WebSockets protocol turned out to be the slowest and the least stable. There are several tests which took much longer than they normally last. gRPC protocol was definitely the fastest in this task. 

Table \ref{tab:mean_fetch_hundred} shows mean test times.

\begin{table}[htbp!]
    \centering
    \caption{Times measured for fetching a hundred elements from database}
    \begin{tabular}{|c|l|r|r|}
        \hline
         \multicolumn{2}{|c|}{\cellcolor[HTML]{EFEFEF}}  & \cellcolor[HTML]{EFEFEF}Docker [$\mu s$] & \cellcolor[HTML]{EFEFEF}Windows [$\mu s$] \\
        \hline
        \cellcolor[HTML]{EFEFEF} & \cellcolor[HTML]{EFEFEF}mean & 24301 & 19728 \\
        \cline{2-4}
        \cellcolor[HTML]{EFEFEF} & \cellcolor[HTML]{EFEFEF}min & 22420 & 17758 \\
        \cline{2-4}
        \cellcolor[HTML]{EFEFEF} & \cellcolor[HTML]{EFEFEF}max & 90189 & 38277 \\
        \cline{2-4}
        \multirow{-4}{*}{\cellcolor[HTML]{EFEFEF}REST} & \cellcolor[HTML]{EFEFEF}$\sigma$ & 3030 & 1600 \\
        \hline
        \cellcolor[HTML]{EFEFEF} & \cellcolor[HTML]{EFEFEF}mean & 11730 & 11611 \\
        \cline{2-4}
        \cellcolor[HTML]{EFEFEF} & \cellcolor[HTML]{EFEFEF}min & 8002 & 9893 \\
        \cline{2-4}
        \cellcolor[HTML]{EFEFEF} & \cellcolor[HTML]{EFEFEF}max & 38763 & 20067 \\
        \cline{2-4}
        \multirow{-4}{*}{\cellcolor[HTML]{EFEFEF}gRPC} & \cellcolor[HTML]{EFEFEF}$\sigma$ & 5403 & 1189 \\
        \hline
        \cellcolor[HTML]{EFEFEF} & \cellcolor[HTML]{EFEFEF}mean & 23107 & 22176 \\
        \cline{2-4}
        \cellcolor[HTML]{EFEFEF} & \cellcolor[HTML]{EFEFEF}min & 20820 & 19266 \\
        \cline{2-4}
        \cellcolor[HTML]{EFEFEF} & \cellcolor[HTML]{EFEFEF}max & 48715 & 44608 \\
        \cline{2-4}
        \multirow{-4}{*}{\cellcolor[HTML]{EFEFEF}GraphQL} & \cellcolor[HTML]{EFEFEF}$\sigma$ & 2567 & 2565 \\
        \hline
        \cellcolor[HTML]{EFEFEF} & \cellcolor[HTML]{EFEFEF}mean & 25327 & 23533 \\
        \cline{2-4}
        \cellcolor[HTML]{EFEFEF} & \cellcolor[HTML]{EFEFEF}min & 22962 & 20888 \\
        \cline{2-4}
        \cellcolor[HTML]{EFEFEF} & \cellcolor[HTML]{EFEFEF}max & 189695 & 51930 \\
        \cline{2-4}
        \multirow{-4}{*}{\cellcolor[HTML]{EFEFEF}WebSockets} & \cellcolor[HTML]{EFEFEF}$\sigma$ & 9444 & 2244 \\
        \hline
    \end{tabular}
    \label{tab:mean_fetch_hundred}
\end{table}


\newpage

\subsection{Memory comparison}

\subsubsection{Insert one element to database}

Table \ref{table:messagememory} presents how many bytes has been transferred in each protocol and operation. The measurements has been collected with WireShark.

\begin{table}[h!]
\caption{Message size in bytes}
\begin{center}
\begin{tabular}{|l|l|l|l|l|lll}
\cline{1-5}
\cellcolor[HTML]{EFEFEF}Operation & \cellcolor[HTML]{EFEFEF}REST & \cellcolor[HTML]{EFEFEF}gRPC & \cellcolor[HTML]{EFEFEF}GraphQL & \cellcolor[HTML]{EFEFEF}WebSockets &  &  &  \\ \cline{1-5}
Fetching element                               & 3536 & 3811 & 3766 & 3469             & & &  \\ \cline{1-5}
Inserting element                               & 3298 & 3769 & 3746 & 3069             & & &  \\ \cline{1-5}
\end{tabular}
\end{center}
\label{table:messagememory}
\end{table}

 Presented table shows that, indisputably, WebSockets offers the best memory usage out of 4 presented communication protocols. In both fetching and inserting it achieved the best result. Other protocols use noticeably more resources, especially gRPC which is the most memory-consuming mechanism of communication in this comparison.

\section{Conclusion}
Out tests showed that gRPC protocol is the fastest in transferring data between client and server. The WebSockets protocol achieved similar results to REST when transferred data was small (inserting and fetching one element). When data was larger (test with a hundred elements) it turned out to be the slowest. REST style was moderately fast. It turned to be slower than gRPC, but was never the worst in any category. GraphQL had some troubles with small data. It was the worst in inserting and fetching one element from database. It was slightly faster than the WebSockets in fetching a hundred elements.

What can easily be visible is that Docker platform is less stable than native OS. There are several tests (especially the WebSockets protocol tests) which show outlier times. This means that Docker might have an impact on a web app performance.

What was also noticed is that the memory usage is in some way associated with protocols performance. WebSockets scored poorly in performance comparison but used the littlest amount of memory to transfer data. The opposite is with gRPC, it has noticed the best performance but was the most memory-consuming at the same time.

Above conclusions and the overall authors feelings lead to another conclusions about proper use cases of each protocol. If a programmer is looking for the fastest way to transfer data and does not care about message size the relevant option will be gRPC. On the other hand, if delivery time is not crucial and there is need of low memory usage the right option will be WebSocket. REST protocol is also an interesting way to communicate. It provides decent memory usage, time performance and what is the most important it is easily accessible, which means that every new client can easily connect to server. There is also GraphQL which provides the worst memory usage to time performance ratio especially in single element operations. This means that this protocol should be used in other way e.g. nested data structures where data is fetched with queries.

\bibliographystyle{fcds_abbrv}
\bibliography{refs}

\begin{thebibliography}{10}

\bibitem{80000_Steam_Games_DataSet}
Deepan.n, 80000 steam games dataset.
\newblock "kaggle.com/datasets/deepann/80000-steam-games-dataset/metadata",
  accessed April, 2022.

\bibitem{grpc_doc}
Grpc official documentation.
\newblock {https://grpc.io/docs/what-is-grpc/core-concepts/}, accessed May 17,
  2022.

\bibitem{proto_doc}
Protocol buffers official documentation.
\newblock {https://developers.google.com/protocol-buffers/docs/overview},
  accessed May 17, 2022.

\bibitem{efficiency_rest_grpc}
Bolanowski~M., Żak~K., Paszkiewicz~A., Ganzha~M., Paprzycki~M., Sowiński~P.,
  Lacalle~I., and Palau~C.~E.
\newblock Eficiency of rest and grpc realizing communication tasks in
  microservice-based ecosystems, 2022.

\bibitem{rest_graphql_convenience}
Brito~G. and Valente~M.~T.
\newblock Rest vs graphql: A controlled experiment.
\newblock In {\em 2020 IEEE international conference on software architecture
  (ICSA)}, pages 81--91. IEEE, 2020.

\bibitem{energy_cost}
Chamas~C.~L., Cordeiro~D., and Eler~M.~M.
\newblock Comparing rest, soap, socket and grpc in computation offloading of
  mobile applications: An energy cost analysis.
\newblock In {\em 2017 IEEE 9th Latin-American Conference on Communications
  (LATINCOM)}, pages 1--6. IEEE, 2017.

\bibitem{performance_graph_rest_soap}
Erlandsson~P. and Remes~J.
\newblock Performance comparison: Between graphql, rest \& soap, 2020.

\bibitem{rfc6455}
Fette~I. and Melnikov~A.
\newblock {The WebSocket Protocol}.
\newblock {RFC} 6455, IETF Trust, December 2011.

\bibitem{http_rfc}
Fielding~R. and Reschke~J.
\newblock Hypertext transfer protocol (http/1.1): Semantics and content.
\newblock Technical report, 2014.

\bibitem{rest_roy}
Fielding~R.~T.
\newblock {\em Architectural styles and the design of network-based software
  architectures}.
\newblock University of California, Irvine, 2000.

\bibitem{frigaard2022graphql}
Frig{\aa}rd~E.
\newblock Graphql vs. rest: A comparison of runtime performance, 2022.

\bibitem{lasocha_badurowicz}
{\L}asocha~W.~P. and Badurowicz~M.
\newblock Comparison of websocket and http protocol performance por{\'o}wnanie
  wydajno{\'s}ci protoko{\l}u websocket i http.
\newblock {\em Journal of Computer Sciences Institute}, 19:67--74, 2021.

\bibitem{inproceedingsSqlNoSqlPerf}
Li~Y. and Manoharan~S.
\newblock A performance comparison of sql and nosql databases.
\newblock pages 15--19, 08 2013.

\bibitem{lubartowicz2020performance}
Lubartowicz~P. and Pa{\'n}czyk~B.
\newblock Performance comparison of web services using symfony, spring, and
  rails examples.
\newblock {\em Journal of Computer Sciences Institute}, 17:384--389, 2020.

\bibitem{makkonen2017performance}
Makkonen~J. et~al.
\newblock Performance and usage comparison between rest and soap web service.
\newblock 2017.

\bibitem{rest_graphql}
Seabra~M., Naz{\'a}rio~M.~F., and Pinto~G.
\newblock Rest or graphql? a performance comparative study.
\newblock In {\em Proceedings of the XIII Brazilian Symposium on Software
  Components, Architectures, and Reuse}, pages 123--132, 2019.

\bibitem{sliwa_panczyk}
{\'S}liwa~M. and Pa{\'n}czyk~B.
\newblock Performance comparison of programming interfaces on the example of
  rest api, graphql and grpc.
\newblock {\em Journal of Computer Sciences Institute}, 21:356--361, 2021.

\end{thebibliography}

\end{document}